\documentclass[preprint,onecolumn,aps,showpacs,showkeys,nofootinbib,superscriptaddress]{revtex4}
\usepackage{amsmath,bbm}
\usepackage{graphicx}

\newcommand{\sla}[1]{/\!\!\!#1}
\newcommand{\calM}{{\cal M}}
\newcommand{\calO}{{\cal O}}

\begin{document}
\title{Narrow-width approximation accuracy}
\date{\today}
\author{C.~F.~Uhlemann}
\email[]{uhlemann@physik.uni-wuerzburg.de}
\affiliation{Institut f\"ur Theoretische Physik, Universit\"{a}t W\"{u}rzburg, D-97074 W\"{u}rzburg, Germany}
\author{N.~Kauer}
\email[]{n.kauer@rhul.ac.uk}
\affiliation{Department of Physics, Royal Holloway, University of London, Egham TW20 0EX, UK}
\begin{abstract}
A study of general properties of the narrow-width approximation (NWA) with 
polarization/spin decorrelation is presented. 
We prove for sufficiently inclusive differential rates of arbitrary 
resonant decay or scattering processes with an 
on-shell intermediate state decaying via a cubic or quartic vertex 
that decorrelation effects vanish and the NWA is 
of order $\Gamma$.
Its accuracy is then determined numerically for all resonant $3$-body decays
involving scalars, spin-$\frac{1}{2}$ fermions or vector bosons.  We specialize
the general results to MSSM benchmark scenarios.  Significant off-shell 
corrections can occur -- similar in size to QCD corrections.  We qualify the 
configurations in which a combined consideration is advisable.  For this purpose, 
we also investigate process-independent methods to improve the NWA.
\end{abstract}
\maketitle
\preprint{}


\section{Introduction\label{intro-section}}

Theoretical arguments and experimental observations indicate that new particles
or interactions play an important role at the TeV scale, which will soon 
become directly accessible at the Large Hadron Collider (LHC).
In the near future we can therefore anticipate ground-breaking discoveries that 
reveal physics beyond the Standard Model (BSM) \cite{bsmhouches}.  
Theoretically appealing extensions of the Standard Model often feature numerous 
additional interacting heavy particles.  BSM phenomenology is hence characterized by 
particle production and cascade decays, which lead to many-particle final states and 
scattering amplitudes with complex resonance structure.
In order to extract the additional Lagrangian parameters of an extended theory 
from collider data, theoretical predictions are required that match the 
experimental accuracies.

The treatment of unstable particles in quantum field theory is complicated
by the absence of asymptotic final states for particles with finite mean
lifetime.  That a consistent theory respecting unitarity, renormalizability and 
causality can be formulated based on external stable states only has been shown
in Ref.~\cite{Veltman:1963th}.  That work, however, does not discuss how to 
carry out fixed-order perturbative calculations involving unstable particles.
Subsequently, several finite-width schemes have been discussed in the literature 
that in principle facilitate cross section calculations for arbitrary processes at 
tree \cite{treeschemes} and loop level \cite{loopschemes}, where the full fixed-order 
scattering amplitude is taken into account.\footnote{%
A method for the inclusion of finite-width effects in event generators 
is discussed in Ref.~\protect\cite{Gigg:2008yc}.}
If the width $\Gamma$ of an unstable particle is much smaller than its mass $M$,
amplitude contributions that feature the corresponding resonant intermediate state 
are generically enhanced by a factor of order $M/\Gamma$.  A theoretically 
consistent method to extract this typically dominant part of the amplitude is 
provided by the narrow-width approximation (NWA) \cite{Pilkuhn}.  The NWA allows to neglect 
nonresonant as well as nonfactorizable 
amplitude contributions, thus leading to significant simplifications for calculations 
in extensions of the Standard Model or calculations of higher order corrections.
When the NWA is applied in such calculations the unstable particle is effectively 
restricted to on-shell states.  We note that the NWA is hence implicitly applied 
whenever branching ratios are extracted from reaction rates.
Recently, it has been observed that the NWA can be unreliable in relevant 
circumstances, namely with decays where a daughter mass approaches the parent mass
or when parton distribution functions are convolved with a resonant
hard scattering process \cite{nwadevs}.  A more detailed study of the accuracy of 
the NWA is thus well motivated.

In Sec.~\ref{nwaprop}, we describe the NWA and clarify the treatment of 
polarization and spin correlations.  We then consider correlation effects for 
 decay and scattering rates in the special case of an on-shell 
intermediate state, which allows us to prove that the NWA error is of order $\Gamma$.
To the best of our knowledge no proof for this result has previously been given in 
the literature.
The polarization/spin decorrelation has been considered in Ref.~\cite{Dicus:1984fu}.
The authors find a result similar to that derived in Sec.~\ref{proofdeco}, but
obtain it by performing the little-group integral for the resonant particle and 
assuming that the subamplitudes transform under some suitable representation of the 
little group.  In our more direct derivation we will explicitly argue for the 
transformation of the subamplitudes and see that the full little-group integral is 
not generally required.

When applied in perturbative calculations, the uncertainty of the NWA is commonly 
estimated as between $\approx\Gamma/M/3$ and $\approx 3\Gamma/M$, 
i.e.~of order $\Gamma/M$ in the physics sense.  In Sec.~\ref{threebodydecays}
we test this assumption systematically for all resonant $3$-body decays
involving scalars, spin-$\frac{1}{2}$ fermions or vector bosons.  We then specialize 
the generic results
to benchmark scenarios in the Minimal Supersymmetric Standard Model (MSSM).
Our findings in Sec.~\ref{threebodydecays} naturally raise the question of
process-independent improvements of the NWA, which we investigate in 
Sec.~\ref{indepimprove}.
We close with a summary in Sec.~\ref{conclusions}.


\section{NWA properties\label{nwaprop}}

\subsection{NWA definition}

Formally, the NWA factorization into production of the unstable particle and its 
subsequent decay is obtained from the full cross section formula by factorizing the 
phase space and integrating out the Breit-Wigner resulting from the squared denominator 
of the resonant particle's propagator.
With $P$ denoting the sum of the incoming particles' momenta and $q$ and $p_i$ 
denoting the momenta of the intermediate resonant particle and the final state 
particles, respectively,
the phase-space factorization of the $n$-particle phase-space element
\begin{equation*}
 d\Phi_n(P;p_1,\dots,p_n)=(2\pi)^4\delta^{(4)}(P-\sum_{i=1}^np_i)\prod_{j=1}^n
\frac{d^3 \vec{p}_j}{(2\pi)^3 2 E_j}
\end{equation*}
into a $j$-particle phase-space element for the production process, 
an $(n-j+1)$-particle phase-space element 
for the decay and an additional $q^2$-integration is given by 
\begin{equation}
 d\Phi_n(P;p_1,\dots,p_n)
=d\Phi_j(P;p_1,\dots,p_{j-1},q) \times \frac{dq^2}{2\pi}\times 
d\Phi_{n-j+1}(q;p_j,\dots,p_n)\,.\label{psfac}
\end{equation}
Denoting the resonant particle's mass and width by $M$ and $\Gamma$, respectively, 
the squared propagator denominator is given by 
$D(q^2):=1/\left[(q^2-M^2)^2+M^2\Gamma^2\right]$.  It is integrated out via
\begin{equation}
\begin{aligned}
\int_{q_\text{min}^2}^{q_\text{max}^2} \frac{dq^2}{2\pi} D(q^2)\sigma_r(q^2) \ &\rightarrow\ 
\int_{-\infty}^{\infty} \frac{dq^2}{2\pi} D(q^2)\sigma_r(q^2)\\ &\rightarrow\ 
\frac{1}{2M\Gamma}\int dq^2\delta(q^2-M^2)\sigma_r(q^2)\ =\ 
\frac{\sigma_r(M^2)}{2M\Gamma}\,,
\end{aligned}\label{Breit-Wigner-integration}
\end{equation}
where $q_\text{min,max}^2$ are the kinematic bounds for $q^2$ and $\sigma_r(q^2)$ denotes 
the residual differential cross section (dependencies on quantities other than $q^2$ 
have been suppressed). 
We note that the second and third expression are asymptotically equal for 
$\Gamma\rightarrow 0$.  For finite $\Gamma$, 
Transformation~(\ref{Breit-Wigner-integration}) constitutes the core NWA.
In the first step one assumes that adding the contributions from $q^2$ regions that 
are not kinematically accessible has a negligible impact, which is the case if 
$M^2$ is not close to $q_\text{min,max}^2$ as measured by $\Gamma$. 
In the second step we assume that the $q^2$-dependence of the remaining integrand 
$\sigma_r(q^2)$, i.e.~the residual squared amplitude and phase-space factors, is 
weak compared to that of the strongly peaked Breit-Wigner.

While the core NWA already provides considerable simplification, it does not 
facilitate complete factorization for nonscalar particles, since the production and 
decay process are linked by correlation effects.
By additionally neglecting polarization/spin correlations between the production and decay parts we obtain the standard NWA.
More specifically, the corresponding numerator of the squared matrix element is 
replaced for unstable vector bosons and fermions, i.e.
\begin{equation}
\begin{aligned}
&|\calM_p^\mu(-g_{\mu\nu}+\frac{q_\mu q_\nu}{M^2})\calM_d^\nu|^2 
=
|\sum_\lambda \calM_p^\mu \epsilon^{\ast\lambda}_{\mu} \epsilon^{\lambda}_{\nu} \calM_d^\nu|^2
\rightarrow
\frac{1}{3}\sum_{\lambda_1,\lambda_2} |\calM_p^\mu \epsilon^{\ast\lambda_1}_{\mu}|^2 |\epsilon^{\lambda_2}_{\nu} \calM_d^\nu|^2,&\\
&|\calM_p(\sla{q}+M)\calM_d|^2
=
|\sum_s\calM_p u_s\bar u_s\calM_d|^2
\rightarrow
\frac{1}{2}\sum_{s_1,s_2}|\calM_p u_{s_1}|^2 |\bar u_{s_2}\calM_d|^2&
\end{aligned}\label{M2-decorrelation}
\end{equation}
with vector boson propagator in unitary gauge and suppressing the argument $q$ for the polarization vectors and spinors.
The resulting factorization of the cross section into $\sigma_p\times\text{BR}$ constitutes the standard NWA factorization.
Two sources of potential problems thus arise: the factorization of the squared amplitude neglects correlation effects, and off-shell effects are ignored when replacing the $q^2$-integration with a constant factor.  We address the former issue in the next section.


\subsection{Polarization/spin decorrelation\label{proofdeco}}

\begin{figure}
\vspace*{0.cm}
\includegraphics[height=3.5cm]{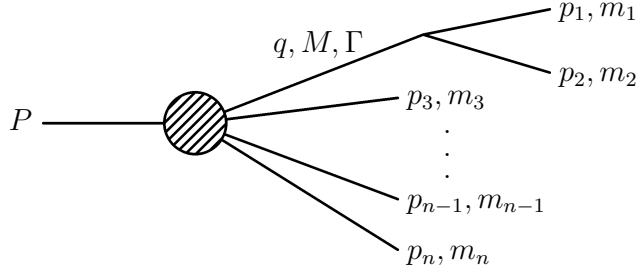}
\caption{\label{polarization-kinematics}
Resonant decay or scattering process kinematics with total incoming momentum $P$.}
\end{figure}

In this section we show that there are no polarization/spin correlation effects for 
total or sufficiently inclusive differential rates\footnote{For a cubic (quartic) decay vertex, 
$\theta_1$ ($\theta_0$, $\phi_0$ and $\phi_1$) has to be integrated out.  The angular variables are defined below Eq.~(\ref{Idef}).} of arbitrary resonant decay or scattering processes with an on-shell 
intermediate state decaying via a cubic or quartic vertex.
We consider processes with an unstable vector boson and note that a  similar
argument applies for unstable fermions. 
The squared amplitude for a process with resonant massive vector boson, which we 
assume to be on mass-shell,  is given by
\begin{equation*}
| \mathcal M_r|^2=|\mathcal M_p^\mu (-g_{\mu \nu}+\frac{q_\mu q_\nu}{ M^2}) 
\mathcal M_d^\nu|^2\,,
\end{equation*}
where the denominator of the squared propagator is suppressed.
By decoupling the polarization sum and inserting a polarization average, i.e.~applying
Transformation~(\ref{M2-decorrelation}),
one obtains the decorrelated squared matrix element
\begin{equation*}
\overline{|\calM_r|^2}
=\left[\mathcal M_p^\mu (-g_{\mu\nu}+\frac{q_\mu q_\nu}{M^2}){\mathcal M_p^{\nu}}^\ast\right] \times \frac{1}{3} \left[\mathcal M_d^\alpha (-g_{\alpha\beta}+\frac{q_\alpha q_\beta}{M^2}){\mathcal M_d^\beta}^\ast\right].
\end{equation*}
As first step, we consider the process of Fig.~\ref{polarization-kinematics}, where the resonant particle is produced in an arbitrary process with total incoming momentum $P$ 
and decays via a $2$-body decay.
To obtain the total rate for a specific process, the matrix elements have to be integrated over the full phase space. Here, we consider the total rate after application of Transformation~(\ref{Breit-Wigner-integration}), which -- up to overall factors -- is given by
\begin{align*}
&\int \prod_{i=1}^n\frac{d^3\vec{p}_i}{2E_i}\; \delta^{(4)}\big(P-\sum_{j=1}^n p_j\big)\,D\!\left(q^2\right) |\mathcal M_r|^2\\
\rightarrow\ &\int \delta^{(4)}(P-q-\sum_{j=3}^n p_j)\, \frac{d^3\vec{q}}{2E_q}\,\prod_{i=3}^n\frac{d^3\vec{p}_i}{2E_i}\,\frac{\delta(q^2-M^2)dq^2}{2M\Gamma}\,\delta^{(4)}\big(q-p_1-p_2\big) \frac{d^3\vec{p}_1}{2E_1}\frac{d^3\vec{p}_2}{2E_2}|\mathcal M_r|^2 \\
=\ \;&\int \frac{\delta^{(4)}(P-q-\sum_{j=3}^n p_j)}{2M\Gamma}\,\frac{d^3\vec{q}}{2E_q} \prod_{i=3}^n\frac{d^3\vec{p}_i}{2E_i}\,  \delta^{(4)}\big(q-p_1-p_2\big)\frac{d^3\vec{p}_1}{2E_1}\frac{d^3\vec{p}_2}{2E_2} |\mathcal M_r|^2 \bigg|_{q^2=M^2}.
\end{align*}
The standard NWA is now obtained by decorrelating the squared amplitude, i.e.~by 
replacing $|\mathcal M_r|^2$ with $\overline{|\calM_r|^2}$.
To determine the effect of this procedure we consider the difference
\begin{equation}
 \int \frac{\delta^{(4)}(P-q-\sum_{j=3}^n p_j)}{2M\Gamma}\,\frac{d^3\vec{q}}{2E_q} \prod_{i=3}^n\frac{d^3\vec{p}_i}{2E_i}\, 
 \underbrace{\delta^{(4)}\big(q-p_1-p_2\big)\frac{d^3\vec{p}_1}{2E_1}\frac{d^3\vec{p}_2}{2E_2} \left(|\mathcal M_r|^2-\overline{|\calM_r|^2}\right)}_{=:H} \bigg|_{q^2=M^2}.\label{Hdef}
\end{equation}
Exploiting that $H$ defined in Expression (\ref{Hdef}) is Lorentz invariant, we choose 
the Gottfried-Jackson frame 
\cite{byckling}, in which $q=(E_q,\vec{0})^T$, to evaluate it.  In this frame our 
assumption $q^2=M^2$ implies $q=(M,\vec{0})^T$.
By integrating out $\vec{p}_2$ and substituting $d^3\vec{p}_1=|\vec{p}_1|^2\; d|\vec{p}_1|\; d\Omega_1$ we find
\begin{align}
H=&\int \frac{|\vec{p}_1|^2\,d|\vec{p}_1|}{2E_1}\;\delta(E_q-E_1-E_2)\;\underbrace{\int d\Omega_1\,\left(|\mathcal M_r|^2-\overline{|\calM_r|^2}\right)}_{=:I}\bigg|_{\textstyle\{\vec{p}_2=-\vec{p}_1,E_q=M\}}.\label{Idef}
\end{align}
We choose $d\Omega_1$ as innermost integration, for which $|\vec{p}_1|$ and $\vec{p}_3,\dots,\vec{p}_n$ are invariables.
Since the production and decay amplitudes $\mathcal M_p^\mu \epsilon^\ast_\mu$ and 
$\mathcal M_d^\nu \epsilon_\nu$ are Lorentz invariant, $\mathcal M_p^\mu$ 
and $\mathcal M_d^\nu$ are Lorentz $4$-vectors.
The spatial component of the decay (production) $4$-vector has to be proportional to 
$\vec{p}_1$ (a function of $\vec{p}_3,\dots,\vec{p}_n$) since no independent other $4$-vectors exist in 
the decay (production) part, i.e.~with $\mathcal M_p=(\mathcal M_p^0,
\vec{\mathcal M}_p)^T$ and $\mathcal M_d=(\mathcal M_d^0,\vec{\mathcal M}_d)^T$ 
we have $\vec{\mathcal M}_d\propto \vec{p}_1$ and $\vec{\mathcal M}_p$ 
fixed.\footnote{The proportionality factor for $\vec{\mathcal M}_d$
can depend on $\vec{p}_i$ only through $|\vec{p}_1|$, which is fixed.}
We therefore choose $d\Omega_1=d\!\cos\!\theta_1\;d\phi_1$ with 
$\theta_1=\angle(\vec{p}_1,\vec{\mathcal M}_p)$.
Furthermore, in the Gottfried-Jackson frame the polarization sum
$-g_{\mu\nu}+q_\mu q_\nu/M^2$ is given by $\operatorname{diag}(0,\mathbbm{1}_3)$ and
for the squared matrix elements we then get 
$$|\mathcal M_r|^2=|\vec{\mathcal M}_p \cdot \vec{\mathcal M}_d|^2=|\vec{\mathcal M}_p|^2 |\vec{\mathcal M}_d|^2\cos^2\theta_1
\qquad \text{and}\qquad
\overline{|\calM_r|^2}=\frac{1}{3}|\vec{\mathcal M}_p|^2 |\vec{\mathcal M}_d|^2\,.$$ 
Evaluating $I$ defined in Expression~(\ref{Idef}), we thus obtain
\begin{align*}
 I &=  
 |\vec{\mathcal M}_p|^2 |\vec{\mathcal M}_d|^2 \int d\!\phi_1\int_{-1}^1 d\!\cos\!\theta_1\ \left(\cos^2\theta_1 - \frac{1}{3}\right)=0\,,
\end{align*}
which proves that polarization decorrelation does not introduce an error for 
on-shell intermediate states in resonant processes of the type shown in 
Fig.~\ref{polarization-kinematics}.

We now consider the case where the unstable 
particle decays via a quartic vertex. 
More specifically, we consider the process of Fig.~\ref{polarization-kinematics} with an additional particle 
with outgoing momentum $p_0$ attached to the decay vertex connecting $q$, $p_1$ and $p_2$.
Again, we consider a total or sufficiently inclusive differential rate 
after application of Transformation~(\ref{Breit-Wigner-integration}) 
and, similar to Eq.~(\ref{Idef}), arrive at
\begin{align*}
H&=\int \frac{d^3\vec{p}_0}{2E_0}\frac{d^3\vec{p}_1}{2E_1}\;\delta(E_q-\sum_{i=0}^2E_i)\left(|\mathcal M_r|^2-\overline{|\calM_r|^2}\right)\bigg|_{\textstyle\{\vec{p}_2=-\vec{p}_0-\vec{p}_1, E_q=M \}}\\
&=\int \frac{|\vec{p}_0|^2\,d|\vec{p}_0|}{2E_0}\;\frac{|\vec{p}_1|^2\,d|\vec{p}_1|}{2E_1}\,d\!\cos\theta_1\,
\delta(E_q-\sum_{i=0}^2E_i)
 \underbrace{\int d\!\cos\theta_0\; d\phi_0\; d\phi_1\,\left(|\mathcal M_r|^2-\overline{|\calM_r|^2}\right)}_{=:J}\bigg|_{\{\dots\}}
\end{align*}
as test for correlation effects.
Now, $|\vec{p}_0|$, $|\vec{p}_1|$, $\theta_1$ and $\vec{\mathcal M}_p$ which, as before, is a function of $p_3,\dots,p_n$ are fixed for 
the inner integrations over $\theta_0$, $\phi_0$ and $\phi_1$.
We parameterized the $\vec{p}_0$-integration as  
$d^3\vec{p}_0=|\vec{p}_0|^2\, d|\vec{p}_0|\; d\!\cos\theta_0\,d\phi_0$, where the orientation of 
$\vec{p}_0$ is measured relative to that of 
$\vec{\mathcal M}_p$, i.e.~$\theta_0=\angle(\vec{p}_0,\vec{\mathcal M}_p)$.
The $\vec{p}_1$-integration was parameterized as $d^3\vec{p}_1=|\vec{p}_1|^2\, 
d|\vec{p}_1|\; d\!\cos\theta_1\,d\phi_1$, oriented such that 
$\theta_1=\angle(\vec{p}_1,\vec{p}_0)$.
$\hat{\vec{r}}_1$, the unit vector in the direction of $\vec{p}_1$, 
is obtained from $\hat{\vec{r}}_0$, the unit vector in the direction of $\vec{p}_0$, 
by first rotating by $\theta_1$ around the axis defined by 
$\hat{\vec{\phi}}_0=(-\sin\phi_0,\cos\phi_0,0)^T$ and then by $\phi_1$ around 
$\hat{\vec{r}}_0$:
\begin{equation}
\vec{p}_1=|\vec{p}_1|\; Q(\hat{\vec{r}}_0,{\phi_1})\; Q(\hat{\vec{\phi}}_0,{\theta_1})\;\hat{\vec{r}}_0\,,
\label{p2-def}
\end{equation}
where 
$
Q(\vec{x},\alpha)_{ij}=-\sum_k \epsilon_{ijk}x_k \sin\alpha + (\delta_{ij}-x_ix_j)\cos\alpha + x_ix_j
$
is the matrix representing a rotation by the angle $\alpha$ 
around the axis defined by the unit vector $\vec{x}=(x_1,x_2,x_3)^T$.  
Since the moduli of $\vec{p}_0$ and $\vec{p}_1$ as well as 
their relative orientation are fixed for the integrations in $J$, the decay matrix element can be 
decomposed as 
\begin{equation}
\vec{\mathcal M}_d=c_0 \;\vec{p}_0+c_1 \;\vec{p}_1+c_2 \;\vec{p}_0\times\vec{p_1}\,,
\label{Mp-decomp}
\end{equation}
where we have exploited that the integration is performed in the Gottfried-Jackson 
frame and $\vec{\mathcal M}_d$ can thus not depend on $p_3,\dots,p_n$.  Note also that $\vec{p}_2=-\vec{p}_0-\vec{p}_1$ is not an independent $3$-vector and hence does not appear in the decomposition.
The coefficients $c_0$, $c_1$ and $c_2$ can only depend on 
$|\vec{p}_0|$, $|\vec{p}_1|$ or 
$\vec{p}_0\cdot\vec{p}_1$ (or equivalently $\theta_1$).  
They are, however, independent of 
the variables of the $J$-integration $\theta_0$, $\phi_0$ and $\phi_1$.
Using the explicit expressions of Eqs.~(\ref{p2-def}) and 
(\ref{Mp-decomp}), $J$ can be evaluated, and we find $J=0$.
We conclude that also no error is introduced by decorrelation when the intermediate particle decays via a quartic vertex.

We now generalize the proof to an on-shell intermediate state decaying via a cubic or 
quartic vertex in an arbitrary resonant process, where the decay products of the 
unstable particle may decay further.
Suppose the particle with momentum $p_1$ decays into $m$ particles with momenta $p_{11},\dots,p_{1m}$.
We map the kinematics of the extended process to the kinematic configuration of 
Fig.~\ref{polarization-kinematics} and factorize the entire phase-space element 
by applying Eq.~(\ref{psfac}) to the intermediate state $p_1$
\begin{equation*}
d\Phi(P;\dots)=d\Phi(P;p_1,p_2,\dots,p_n)\;\frac{dp_1^2}{2\pi}\;d\Phi(p_1;p_{11},\dots,p_{1m})\,.
\end{equation*}
By first integrating $d\Phi(P;p_1,p_2,\dots,p_n)$, we construct the situation that we dealt with before, the only difference being that in the decomposition of the matrix element $\mathcal M_d$ we now have additional terms involving $\vec{p}_{1i}$, $i=1,\dots,m$.
However, in the crucial inner integrations of $d\Phi(P;p_1,p_2,\dots,p_n)$ the 
relative orientations of these momenta are fixed and thus they transform like 
$\vec{p}_1$ under rotations.
The absence of on-shell correlation effects can therefore be shown in the same way 
as above.
Generalization to the case where multiple decay products decay further or where 
correlation effects for multiple on-shell intermediate states are considered 
is straightforward.
Note that this extension again holds not only for total rates, but also for 
sufficiently inclusive differential rates.


\subsection{Order of the NWA error\label{prooforder}}

We can now prove that the relative error 
\begin{equation}
R=\frac{\Gamma_\text{OFS}-\Gamma_\text{NWA}}{\Gamma_\text{NWA}}
\label{Rdef}
\end{equation}
is of $\mathcal O  (\Gamma)$ for the decay and scattering rates considered in 
Sec.~\ref{proofdeco}.\footnote{%
We illustrate the proof using total decay rates, but note that it also applies to sufficiently inclusive differential decay and scattering rates as considered in Sec.~\ref{proofdeco}.}
A priori this is not clear. The error as function of $\Gamma$ could, 
for example, behave like $\Gamma^\kappa$ with $0< \kappa <1$.  In this case, a
series expansion around $\Gamma = 0$ would be invalid.
We consider the off-shell decay rate
\begin{equation*}
\Gamma_\text{OFS} = \int_{q_\text{min}^2}^{q_\text{max}^2} \frac{d q^2}{2 \pi} D(q^2,\Gamma)\,
\Gamma_r(q^2)\,,
\end{equation*}
where kinematic limits define the integration bounds $q_\text{min,max}^2$ and 
we have explicitly noted the dependence of the squared propagator denominator 
on $\Gamma$.
$\Gamma_r(q^2)$ denotes the residual, $\Gamma$-independent integrand.\footnote{%
Since interference effects with nonresonant diagrams are not resonance enhanced, 
we do not consider them here.}
More specifically, $\Gamma_r$ is the squared combination of the production matrix element, the numerator of the propagator and the decay matrix element, integrated over $d\Phi_p$ and $d\Phi_d$ and includes overall factors like $1/(2\sqrt{P^2})$.
The decay rate in NWA is given by
$\Gamma_\text{NWA}=\Gamma_p(M^2)\times\Gamma_d(M^2)/\Gamma$
where $\Gamma_p$ and $\Gamma_d$ are the production rate and partial decay rate, respectively.
We can now write:
\begin{eqnarray*} 
\Gamma_\text{OFS}-\Gamma_\text{NWA} &=& 
\int_{q_\text{min}^2}^{q_\text{max}^2} \frac{d q^2}{2 \pi} D(q^2,\Gamma)\Gamma_r(q^2) - \frac{\Gamma_p(M^2)\Gamma_d(M^2)}{ \Gamma}\nonumber\\
&=& \int_{q_\text{min}^2}^{q_\text{max}^2} \frac{d q^2}{2 \pi} D(q^2,\Gamma)\left(\Gamma_r(q^2)-\Gamma_p(M^2)2M \Gamma_d(M^2)\right) - \alpha(\Gamma)\, \frac{\Gamma_p(M^2) \Gamma_d(M^2)}{ \Gamma}\nonumber\\
&\overset{(\ast)}{=}& \int_{q_\text{min}^2}^{q_\text{max}^2} \frac{d q^2}{2 \pi} D(q^2,\Gamma)\left( \Gamma_r(q^2)-\Gamma_r(M^2)\right) - \alpha(\Gamma)\, \frac{\Gamma_p(M^2) \Gamma_d(M^2)}{ \Gamma}
\end{eqnarray*}
with
$\alpha:=1 - 2 M \Gamma \int_{q_\text{min}^2}^{q_\text{max}^2} D(q^2,\Gamma) d q^2/(2 \pi)$.
Note that $\alpha=0$ for $q_\text{min,max}^2=\pm\infty$.
As shown in Sec.~\ref{proofdeco}, decorrelation effects vanish and $\Gamma_r(M^2)$ factorizes, which has been used in ($\ast$).
For the relative deviation $R$ we get
\begin{eqnarray*}
|R|&=&\bigg |\frac{\Gamma_\text{OFS}-\Gamma_\text{NWA}}{\Gamma_\text{NWA}}\bigg |
\leq\bigg |\int_{q_\text{min}^2}^{q_\text{max}^2} \frac{d q^2}{2 \pi} D(q^2,\Gamma)\,\Gamma\,\frac{\Gamma_r(q^2)-\Gamma_r(M^2)}{\Gamma_p (M^2)\Gamma_d(M^2)}\bigg |+|\alpha(\Gamma)|\,.
\end{eqnarray*}
Assume now that $\Gamma_r(q^2)$ is twice continuously differentiable in the 
kinematically allowed phase-space region and that
$\Gamma_p(q^2)\Gamma_d(q^2)=0 \Leftrightarrow q^2 \in \{q_\text{min}^2,q_\text{max}^2\}$.
Then, the function
\begin{equation}
g_M(q^2):=\frac{\Gamma_r (q^2)}{\Gamma_p(M^2)\Gamma_d(M^2)}
\end{equation}
is twice continuously differentiable with respect to $q^2$ for all $q^2\in [q_\text{min}^2,q_\text{max}^2]$ and $M^2\in(q_\text{min}^2,q_\text{max}^2)$, which we assume in the following, since
we consider resonant intermediate states.
By second order Taylor expansion we then get
\begin{equation*}
g_M(q^2)=g_M(M^2)+g_M^\prime(M^2)\cdot (q^2-M^2) + \frac{1}{2}g_M^{\prime\prime}(x)\cdot (q^2-M^2)^2
\end{equation*}
with $x\in [q_\text{min}^2,q_\text{max}^2]$.
Note that $x$ depends on $q^2$.  As continuous functions on the compact set 
$[q_\text{min}^2,q_\text{max}^2]$, $g_M^{\prime}$ and $g_M^{\prime\prime}$ are then 
bounded and we can find $L_1>0$ and $L_2>0$ such that
\begin{equation}
|g_M^{\prime}(q^2)|\leq L_1\quad\text{and}\quad |g_M^{\prime\prime}(q^2)|\leq L_2\quad\forall q^2\in[q_\text{min}^2,q_\text{max}^2]\,.
\end{equation}
We can now further evaluate $|R|$ to obtain
\begin{eqnarray*}
|R|&\leq&\bigg |\int_{q_\text{min}^2}^{q_\text{max}^2} \frac{d q^2}{2 \pi} D(q^2,\Gamma)\,\Gamma\,\left(g_M(q^2)-g_M(M^2)\right)\bigg |+|\alpha(\Gamma)|\nonumber\\
&=&\frac{\Gamma}{2\pi}\bigg |\int_{q_\text{min}^2}^{q_\text{max}^2} d q^2 D(q^2,\Gamma)\left(g_M^\prime(M^2)\cdot (q^2-M^2) + \frac{1}{2}g_M^{\prime\prime}(x)\cdot (q^2-M^2)^2\right)  \bigg |+|\alpha(\Gamma)|\nonumber\\
&\leq& \frac{\Gamma L_1}{2\pi}\bigg |\int_{q_\text{min}^2}^{q_\text{max}^2} d q^2 \frac{q^2-M^2}{(q^2-M^2)^2+M^2 \Gamma^2}  \bigg |
+\frac{\Gamma L_2}{4 \pi}\int_{q_\text{min}^2}^{q_\text{max}^2} d q^2 \bigg |\frac{(q^2-M^2)^2}{(q^2-M^2)^2+M^2 \Gamma^2}  \bigg |+|\alpha(\Gamma)|\nonumber\\
&=&\frac{\Gamma L_1}{4\pi}\bigg|  \log\frac{(q_\text{max}^2-M^2)^2+M^2\Gamma^2}{(q_\text{min}^2-M^2)^2+M^2\Gamma^2}     \bigg|\\
&&+\frac{\Gamma L_2}{4 \pi} \bigg | q_\text{max}^2+M\Gamma \arctan \frac{M^2-q_\text{max}^2}{M\Gamma}-q_\text{min}^2-M\Gamma \arctan \frac{M^2-q_\text{min}^2}{M\Gamma}\bigg |+|\alpha(\Gamma)|\,. 
\end{eqnarray*}
By employing the Taylor expansions
\begin{align*}
\log \frac{k+x^2}{l+x^2} &= \log \frac{k}{l} + \mathcal O (x^2)\,,\\
\arctan\frac{k}{x} &= \operatorname{sign}(k) \frac{\pi}{2}-\frac{x}{k} + \mathcal O (x^3)
\end{align*}
for small positive $x$ to expand the log and arctan functions for small $\Gamma$, 
we get
\begin{eqnarray*}
|R|&\leq& \frac{\Gamma L_1}{4\pi} \bigg|\log \frac{(q_\text{max}^2-M^2)^2}{(M^2-q_\text{min}^2)^2}\bigg|  + |\alpha(\Gamma)|
+\frac{\Gamma L_2}{4 \pi} ( q_\text{max}^2-q_\text{min}^2+\pi M\Gamma)+ \mathcal O \big(\Gamma^2\big).
\end{eqnarray*}
Lastly, we evaluate and expand $\alpha$:
\begin{eqnarray*}
\alpha &=& 1 - 2 M \Gamma \int_{q_\text{min}^2}^{q_\text{max}^2} \frac{d q^2}{2 \pi} D(q^2,\Gamma)
= 1 - \frac{1}{\pi} \left(  \arctan\frac{q_\text{max}^2-M^2}{M \Gamma} - \arctan\frac{q_\text{min}^2-M^2}{M \Gamma}  \right)\\
&=&\frac{M\Gamma}{\pi}\frac{(q_\text{max}^2-q_\text{min}^2)}{(q_\text{max}^2-M^2)(M^2-q_\text{min}^2)} + \mathcal O \big(\Gamma^3\big)\,,
\end{eqnarray*}
and obtain as final result:
\begin{gather}
\begin{split}
\frac{|R|}{\Gamma}\ \leq\ \, & \frac{L_1}{4\pi} \bigg|\log \frac{(q_\text{max}^2-M^2)^2}{(q_\text{min}^2-M^2)^2}\bigg|
+\frac{L_2}{4 \pi} \left( q_\text{max}^2-q_\text{min}^2\right)\\
&+\frac{M}{\pi}\frac{(q_\text{max}^2-q_\text{min}^2)}{(q_\text{max}^2-M^2)(M^2-q_\text{min}^2)}+ \mathcal O \big(\Gamma\big)\,,
\end{split}\label{R-bound}
\end{gather}
where $L_1$ and $L_2$ are $\Gamma$-independent constants with mass dimension $-1$ and $-3$, respectively.
We have thus shown that $R$ is of $\mathcal O (\Gamma)$ if $M$ is in the interior of the kinematically allowed region.  $R/\Gamma$ is then finite in the limit 
$\Gamma\rightarrow 0$. 
Note that this is a stronger statement than the asymptotic equality of 
$\Gamma_\text{OFS}$ and $\Gamma_\text{NWA}$ for $\Gamma\rightarrow 0$, which immediately follows 
from the asymptotic equality of the off-shell and core NWA expressions 
(see Transformation~(\ref{Breit-Wigner-integration})) and the absence of on-shell 
correlation effects.
$R=\mathcal O (\Gamma)$ does, however, not guarantee that $|R|\approx \Gamma/M$
as suggested by the scales that occur in the unstable particle propagator.
The leading terms in $\Gamma$ on the right-hand side of Ineq.~(\ref{R-bound}), while $\Gamma$-independent, may 
nevertheless become arbitrarily large as $M$ approaches the kinematic limits.
This can be seen directly by inspecting the factors that are explicitly 
$(q_\text{min,max}^2-M^2)$ dependent, but is also due to the fact that $L_1$ and $L_2$ may 
become large as $M^2$ approaches $q_\text{min}^2$ or $q_\text{max}^2$.


\section{NWA accuracy for resonant three-body decays\label{threebodydecays}}

In Sec.~\ref{nwaprop} we have already determined several necessary conditions for a 
small relative NWA error $|R|$:  The standard NWA will only be accurate if 
polarization/spin correlation effects can be neglected.  Furthermore, 
Transformation (\ref{Breit-Wigner-integration}) shows that 
$|(q^2_\text{min,max})^{1/2}-M|\gtrsim\Gamma$ is necessary to effectively 
eliminate the dependence on the $q^2$-integration bounds.  In the next section
we systematically explore NWA deviations for the process class that
allows application of the NWA while featuring minimal complexity 
in order to identify other accuracy limiting factors.


\subsection{Model-independent analysis\label{modindep}}

\begin{figure}
\vspace*{0.cm}
\includegraphics[height=3.5cm]{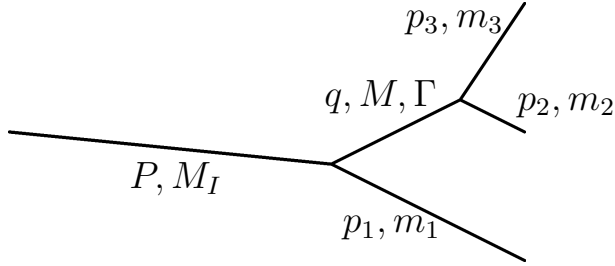}
\caption{\label{threebody-kinematics}
Resonant $3$-body decay kinematics.}
\end{figure}

\begin{figure}
\vspace*{0.cm}
\begin{minipage}[c]{.49\linewidth}
\flushleft \includegraphics[height=6.7cm]{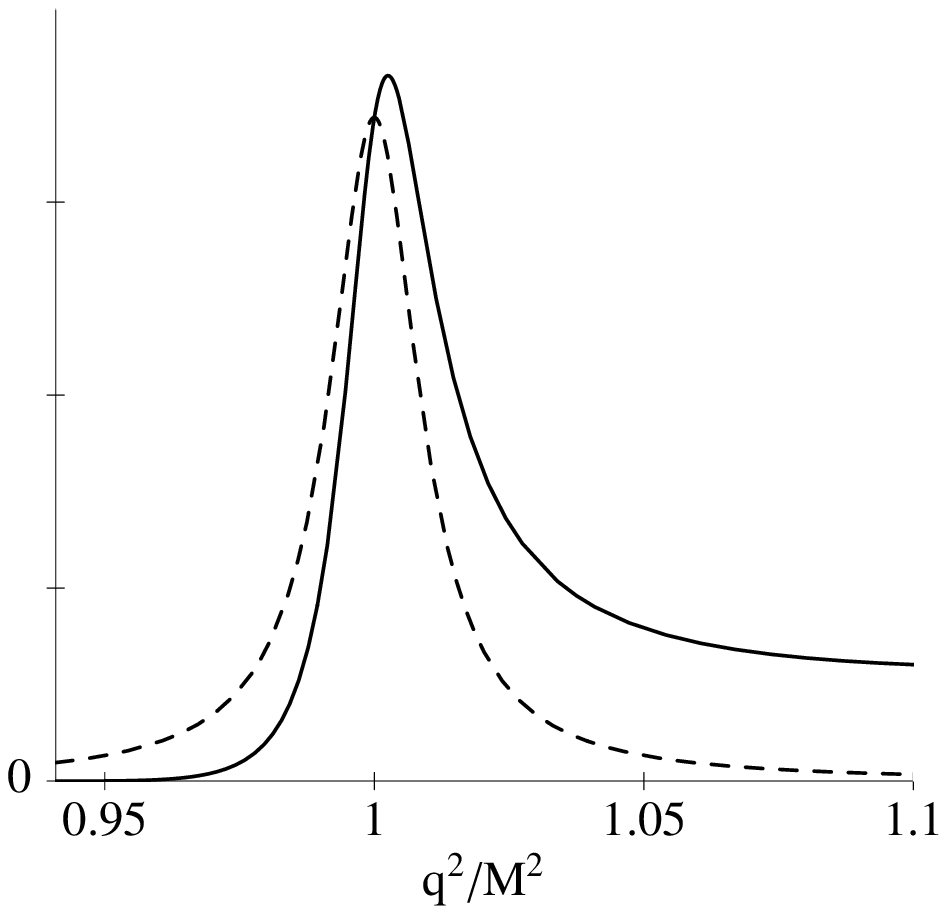}
\end{minipage} \hfill
\begin{minipage}[c]{.49\linewidth}
\flushright \includegraphics[height=6.7cm]{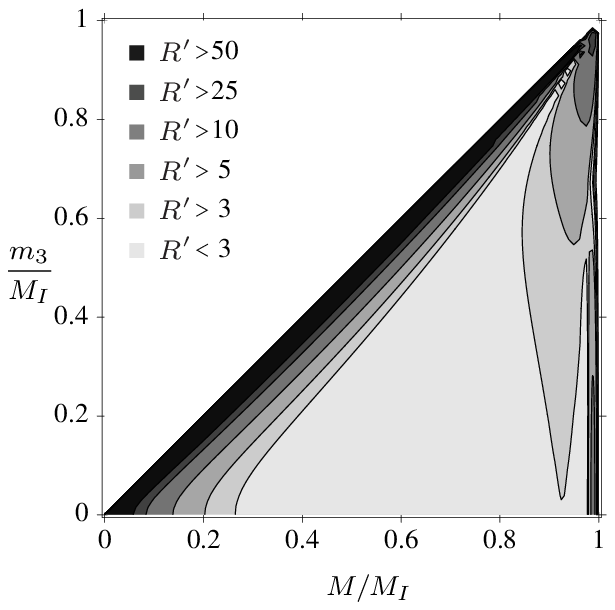}
\end{minipage}\\[0.2cm]
\caption{
Resonant $1\to 3$ decay SSS-SSV (see main text):
The graph displays the 
$q^2$-dependence of the Breit-Wigner that is integrated out 
in the NWA (dashed) and of the complete integrand of Eq.~(\protect\ref{ofsdec_eq:ampfac}) (solid) 
for $m_3=M-3\Gamma$.  The contour plot shows $R':=|R|/(\Gamma/M)$, 
i.e.~the magnitude of the 
relative NWA error in units of $\Gamma/M$, as function of $m_3$ and $M$ 
with $m_1=m_2=0$.  The width-mass ratio $\Gamma/M$ is $0.01$.
\label{ofsdec_fig:sss-ssv}}
\end{figure}

\begin{figure}
\includegraphics[height=6.7cm]{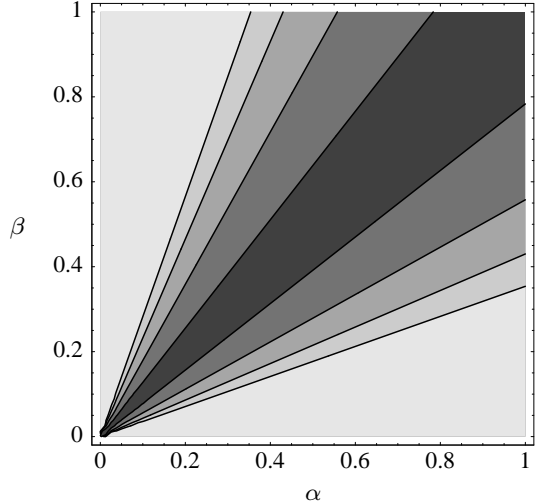}
\caption{\label{ffv-vff}
Dependence of the magnitude of the relative NWA error $R'$ 
(see Fig.~\protect\ref{ofsdec_fig:sss-ssv})
on the strength of the chiral components 
of the 1st-stage decay vertex $\gamma^\mu(\alpha P_L+\beta P_R)$ 
for process FFV-VFF at the parameter point 
$M/M_I=0.68$, $m_1/M_I=0.3$ and $m_2=m_3=0$ with $\Gamma/M=0.01$.
Color code as in Fig.~\protect\ref{ofsdec_fig:sss-ssv}.}
\end{figure}

We have systematically probed the NWA accuracy for resonant $3$-body decays
involving scalars, spin-$\frac{1}{2}$ fermions or vector bosons.  
By inserting these particles into the topology of Fig.~\ref{threebody-kinematics}
one obtains 48 generic decay processes.  
A particular process is selected by giving type (4-momentum, mass) for each particle:
\begin{equation*}
 T_I (P_I, M_I) \to T_1 (p_1, m_1),\;  T (q, M)\ \ \text{and}\ \ T (q, M) \to  T_2 (p_2, m_2),\;  T_3 (p_3, m_3),
\end{equation*}
where type can be scalar (S), fermion (F) or vector boson (V).
The width of the intermediate particle with momentum $q$ is $\Gamma$.  
We have created a custom, model-independent program that allows to calculate 
total decay rates with and without NWA for all 48 processes.  
The program has been validated using \texttt{SMadGraph} \cite{Cho:2006sx} and 
\texttt{SDECAY} \cite{Muhlleitner:2003vg}.\footnote{%
Unitary gauge has been applied throughout.  The employed Feynman rules have been 
double-checked using Ref.~\cite{mssmfeynmanrules}.}
Using this program, we scanned the parameter space for large values of $|R|$ 
by varying masses, width and couplings.  Note that coupling constants typically 
cancel in $R$ with the exception of the relative strength of the chiral components 
of SFF and VFF vertices.  This dependence has been taken into account in the scan.
Qualitatively, we find that the NWA error will exceed order $\Gamma/M$ 
for mass configurations in an extended vicinity of kinematic bounds, but not 
for sufficiently central configurations.  A detailed description and discussion 
of the scan and its results can be found in Ref.~\cite{Uhlemann:diplomarbeit}.  
Here, we choose the process that features the largest deviations, SSS-SSV, to 
explain the mechanism that can lead to large deviations even when the 
$q^2$-integral is not noticeably cut off by kinematic bounds.
The largest deviations occur when all but one final state mass are small.  We therefore
set $m_1=m_2=0$, but keep $m_3\neq 0$.  The $q^2$-integrand for this process is then 
given by
\begin{equation}
\label{ofsdec_eq:ampfac}
\left(1-\frac{q^2}{M_I^2}\right) \left(1-\frac{m_3^2}{q^2}\right)
\left(\frac{(q^2-m_3^2)^2}{m_3^2}\right)\frac{1}{(q^2-M^2)^2+M^2 \Gamma^2}\,,
\end{equation}
where the 1st- and 2nd-stage decay phase-space elements contribute the first and 
second factor, respectively, and the 2nd-stage decay matrix element gives the 
third factor.
When $m_3^2 \lesssim M^2$ the second and third factor effect a strong 
deformation of the Breit-Wigner shape, which, together with the resulting large 
NWA deviations, is displayed in Fig.~\ref{ofsdec_fig:sss-ssv}.
The deviation grows with
increasing power of the deforming factors.
When $M$ approaches the lower kinematic bound, $|R|$ 
is sensitive to the type of the 2nd-stage decay, 
which determines the power of the factor that 
deforms the Breit-Wigner peak.  While this factor enhances the Breit-Wigner 
tail, the factor of the 1st-stage decay suppresses it.
And vice versa for the upper bound.
Generally, we find stronger effects for SSV, VSV, FFV, VVV and SVV than for 
FSF, SFF, VFF, VSS and SSS vertices.
As mentioned above, coupling parameters do not affect the relative NWA error 
if they factorize.
However, for processes with SFF or VFF vertices, a strong dependence 
on the relative strength of the chiral components can exist.  In Fig.~\ref{ffv-vff},
we demonstrate this for the FFV-VFF decay.
We conclude this section with a caveat: $\Gamma_\text{OFS}$ has been calculated including 
the full range of invariant masses $q^2$ for the intermediate resonant state.
In a model-independent analysis we could, however, not take into account 
interference with amplitude contributions with different resonance structure
or with nonresonant diagrams.
In gauge theories, such contributions can in principle be important.
In the next section we will consider examples in specific models.


\subsection{MSSM analysis \label{results-section}}
We now specialize our model-independent analysis to benchmark scenarios in the MSSM.
We determine for each $1\to 3$ decay\footnote{%
We consider final states with stable particles and up to one unstable 
particle.  This is of interest, since subsequent decays do not generally mitigate 
the deviations \cite{nwadevs}.}
in the MSSM and each benchmark parameter point 
of the Snowmass Points and Slopes (SPS) \cite{Allanach:2002nj} whether the resulting 
process is resonant, in which case the NWA accuracy $R$ of Eq.~(\ref{Rdef}) is 
calculated.  
The SPS mass spectra and decay widths have been generated using 
\texttt{SOFTSUSY} \cite{Allanach:2001kg} and \texttt{SDECAY}, respectively.  
As example, we select the most thoroughly studied point SPS 1a (defined by the 
parameter set $M_0=100~{\rm GeV}$, $M_{1/2}=250~{\rm GeV}$, $A_0=-100~{\rm GeV}$, 
$\tan\beta=10$, $\mu>0$) and display in Table~\ref{table-sps1} resonant 3-body decay modes 
with $|R|>5\,\Gamma/M$ for which the distance between $M$ and 
kinematic bounds is larger than $5$ widths.  Note that for 
$\tilde b_1\rightarrow W^- \tilde t_1\rightarrow W^- b \widetilde \chi_2^+$
the NWA error $R$ is negative.  All other processes in Table \ref{table-sps1}
have positive $R$.
A complete listing of results with $|R|>5\,\Gamma/M$ can be found in 
Ref.~\cite{Uhlemann:diplomarbeit}.
\begin{table}
\vspace*{0.cm}
\caption{\label{table-sps1}
Resonant 3-body decay modes in the MSSM at SPS 1a, for which $R^\prime$, 
i.e.~the magnitude of the relative NWA error in units of $\Gamma/M$
(see Fig.~\protect\ref{ofsdec_fig:sss-ssv}) is larger than $5$, and 
$\Delta:=\min\left(M_I-m_1-M, M-m_2-m_3\right)/\Gamma$, i.e.~the 
minimal distance between $M$ and the kinematic bounds $(q_\text{min,max}^2)^{1/2}$ 
in units of $\Gamma$, is larger than $2$.
}
\vspace*{0.5cm}
\renewcommand{\tabcolsep}{0.4cm}
\begin{tabular}{|c|c|c|c|c|c|}
\hline
process &  $R^\prime$ &\ $\Gamma/M$ $[\%]$ & \ $\Delta$\ \\
\hline
  $H^+\rightarrow \tilde \tau_1^* \tilde\nu_{\tau}\rightarrow \tilde \tau_1^* \tau \widetilde \chi_1^+$ & $284$ & $0.080$ & $8.4 $ \\[1mm]
  $\widetilde \chi_1^+\rightarrow \widetilde \chi_1^0 W^+\rightarrow \widetilde \chi_1^0 e^+ \nu_e$ & $5.21$ & $2.5$ & $2.6 $ \\[1mm]
  $\widetilde \chi_1^+\rightarrow \widetilde \chi_1^0 W^+\rightarrow \widetilde \chi_1^0 u \bar{d}$ & $5.21$ & $2.5$ & $2.6 $ \\[1mm]
  $\widetilde \chi_2^+\rightarrow e^+ \tilde\nu_e\rightarrow e^+ e^- \widetilde \chi_1^+$ & $180$ & $0.081$ & $24 $ \\[1mm]
  $\widetilde \chi_3^0\rightarrow \bar{\nu}_e \tilde\nu_e\rightarrow \bar{\nu}_e \nu_e \widetilde \chi_2^0$ & $125$ & $0.081$ & $28 $ \\[1mm]
  $\widetilde \chi_3^0\rightarrow \bar{\nu}_e \tilde\nu_e\rightarrow \bar{\nu}_e e^- \widetilde \chi_1^+$ & $168$ & $0.081$ & $24 $ \\[1mm]
  $\widetilde \chi_4^0\rightarrow \nu_e \tilde\nu_e\rightarrow \nu_e \nu_e \widetilde \chi_2^0$ & $135$ & $0.081$ & $28 $ \\
  $\widetilde \chi_4^0\rightarrow \nu_e \tilde\nu_e\rightarrow \nu_e e \widetilde \chi_1^+$ & $181$ & $0.081$ & $24 $ \\
  $\tilde b_1\rightarrow W^- \tilde t_1\rightarrow W^- b \widetilde \chi_2^+$ & $10.1$ & $0.51$ & $7.7 $ \\[1mm]
$\tilde g\rightarrow d \tilde d_L^*\rightarrow d \bar{d} \widetilde \chi_1^0$ & $9.54$ & $0.94$ & $7.4 $ \\[1mm]
  $\tilde g\rightarrow d \tilde d_L^*\rightarrow d \bar{d}  \widetilde \chi_2^0$ & $7.89$ & $0.94$ & $7.4 $ \\[1mm]
  $\tilde t_1\rightarrow b \widetilde \chi_2^+\rightarrow b \tilde e_L^* \nu_e$ & $25.4$ & $0.66$ & $6.2 $ \\[1mm]
  $\tilde t_1\rightarrow b \widetilde \chi_2^+\rightarrow b \tilde\nu_e e^+$ & $28.1$ & $0.66$ & $6.2 $ \\[1mm]
  $\tilde t_1\rightarrow b \widetilde \chi_2^+\rightarrow b \tilde\nu_{\tau} \tau^+$ & $25.7$ & $0.66$ & $6.2 $ \\[1mm]
\hline
\end{tabular}
\end{table}

\begin{figure}
\vspace*{0.cm}
\includegraphics[height=5.5cm]{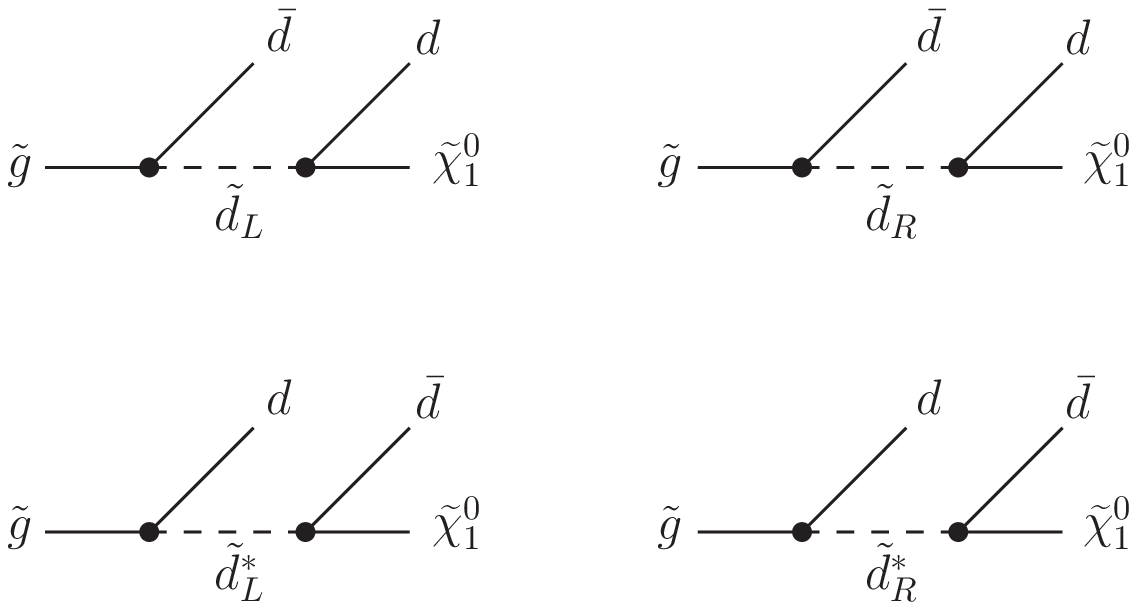}
\caption{\label{go_n1ddb}
Complete set of Feynman graphs for the MSSM $3$-body decay 
$\tilde g \rightarrow \widetilde \chi_1^0 d \bar{d}$.}
\end{figure}

The tabulated results are intended as guideline to alert the NWA user of potential 
large errors.  In such cases a more detailed on-/off-shell comparison including 
all amplitude contributions is called for.  We exemplify this by discussing two
specific cases.  The first process is 
$\tilde g \rightarrow \widetilde \chi_1^0 d \bar{d}$,
which proceeds via intermediate $\tilde d_{L,R}$ states (see 
Fig.~\ref{go_n1ddb}).
At SPS 1a, one has 
$M_{\tilde g}=607.7~{\rm GeV}$, $M_{\tilde d_R}=545.2~{\rm GeV}$, 
$M_{\tilde d_L}=568.4~{\rm GeV}$ and $M_{\widetilde \chi_1^0}=96.7~{\rm GeV}$,
which implies that all intermediate states can be on-shell.
As seen in Table \ref{table-sps1}, one finds $R\approx 10\,\Gamma/M$ for the 
graph with intermediate $\tilde d_L$.\footnote{%
Note the equality of the total decay rates with intermediate $\tilde d_{L,R}$ and 
$\tilde d_{L,R}^\ast$ state due to the $C$ invariance of $|\calM|^2$.}
In contrast, the graph with intermediate $\tilde d_R$ is not affected by an 
unexpectedly large NWA error ($R\approx 2\,\Gamma/M$).  This can be traced to the 
fact that $\tilde d_L$ has a much higher decay probability than $\tilde d_R$ 
due to the chiral structure of the chargino and neutralino couplings.
Consequently, the total width over mass ratio is about $1\%$ for $\tilde d_L$, 
but only about $0.05\%$ for $\tilde d_R$, i.e.~in the latter case 
$|(q^2_\text{min,max})^{1/2}-M|> 200\Gamma$ and hence the absence of large NWA errors.
When also taking into consideration the much larger resonant enhancement  
for the $\tilde d_R$ in comparison with the $\tilde d_L$ mediated process, 
one can expect that the former process in NWA provides a good approximation
to the full $\tilde g \rightarrow \widetilde \chi_1^0 d \bar{d}$ rate.
When calculated in NWA, the decay rate of the $\tilde d_L$ mediated process 
is indeed approximately $100$ times smaller than the decay rate of the 
$\tilde d_R$ mediated process.
Under these circumstances interference effects are likely to provide the leading 
correction to the NWA estimate.
Interference between amplitudes with intermediate states related by 
charge conjugation is Breit-Wigner suppressed, because almost always either the 
invariant mass 
of $\widetilde \chi_1^0 d$ or $\widetilde \chi_1^0 \bar{d}$ will be close to $M$, 
i.e.~only one of the propagators will be resonant.
Furthermore, interference between amplitudes with $\tilde d_L$ and $\tilde d_R$ 
intermediate states is highly suppressed, since the overlap of the 
corresponding resonances is negligible:
$M_{\tilde d_L}-M_{\tilde d_R}\gg \Gamma_{\tilde d_L}+\Gamma_{\tilde d_R}$.
We computed the total decay rate for 
$\tilde g \rightarrow \widetilde \chi_1^0 d \bar{d}$ 
by Monte-Carlo integration of the \texttt{SMadGraph}-generated complete matrix element.
Comparing this off-shell result to the NWA result for the $\tilde d_R$ 
contribution, we find a relative deviation of $1.2\%$, which confirms our expectation.

\begin{figure}
\vspace*{0.cm}
\includegraphics[height=2.5cm]{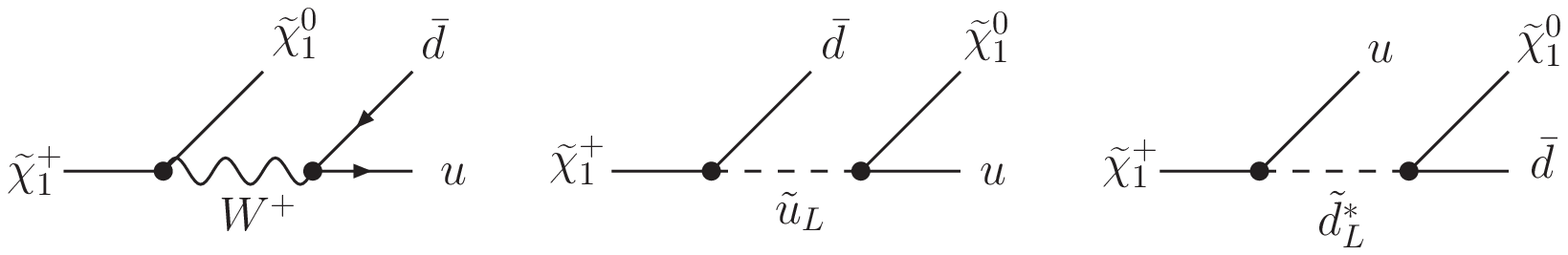}
\caption{\label{x1p_n1udb}
Complete set of Feynman graphs for the MSSM $3$-body decay 
$\widetilde \chi_1^+\rightarrow \widetilde \chi_1^0 u \bar{d}$.}
\end{figure}

As second example, we consider the process 
$\widetilde \chi_1^+\rightarrow \widetilde \chi_1^0 u \bar{d}$,
which proceeds via intermediate $W^+$, $\tilde u_L$ and $\tilde d_L^\ast$ states (see 
Fig.~\ref{x1p_n1udb}).
At SPS 1a, one has 
$M_{\widetilde \chi_1^+}=181.7~{\rm GeV}$, $M_{\widetilde \chi_1^0}=96.7~{\rm GeV}$, 
$M_{W}=79.8~{\rm GeV}$, $M_{\tilde u_L}=561.1~{\rm GeV}$ and 
$M_{\tilde d_L}=568.4~{\rm GeV}$, which implies that only the 
intermediate $W^+$ state can be on-shell.
For the 1st-stage decay $\widetilde \chi_1^+\rightarrow \widetilde \chi_1^0 W^+$ 
with\footnote{At SPS1a' \cite{AguilarSaavedra:2005pw} the branching ratio is $1.3\%$.} 
$\text{BR}=7.5\%$
the minimal distance to the kinematic bounds is $M_I-m_1-M=2.6\Gamma$,
which is not large enough to rule out an unexpectedly large NWA error.  
In fact, Table \ref{table-sps1} shows that $R=13\%=5.2\,\Gamma/M$ for $\widetilde 
\chi_1^+\rightarrow \widetilde \chi_1^0 W^+\rightarrow \widetilde \chi_1^0 u \bar{d}$.
Primarily via interference the nonresonant $\tilde u_L$ and $\tilde d_L^\ast$
contributions slightly reduce the NWA error to $R=11\%$, where
the total decay rate for $\widetilde \chi_1^+\rightarrow \widetilde \chi_1^0 u \bar{d}$ with complete matrix elements is compared to the $\widetilde 
\chi_1^+\rightarrow \widetilde \chi_1^0 W^+\rightarrow \widetilde \chi_1^0 u \bar{d}$
rate in NWA.  Since the 1st-decay stage is not affected by QCD corrections, 
this NWA error is particularly significant.


\section{Process-independent NWA improvement\label{indepimprove}}

In this section we propose a modified NWA prescription that preserves the 
simplifying NWA features and improves the behavior close to kinematic bounds
by taking into account phase-space properties.
The standard NWA is obtained by replacing the distribution 
$\psi: f\mapsto \int \frac{dq^2}{2\pi} D(q^2) f(q^2)$, 
which -- acting on the residual differential cross section $\sigma_r$ -- 
gives the total cross section\footnote{Application to cross sections 
and decay rates is analogous.},
by a rescaled Dirac distribution, i.e.
\begin{equation}
\psi(\sigma_r)=\int \frac{dq^2}{2\pi} D(q^2)\sigma_r(q^2)\quad\rightarrow\quad 
\psi_\text{\,NWA}(\sigma_r)=\frac{1}{2M\Gamma}\int dq^2\delta(q^2-M^2)\sigma_r(q^2)\,,
\label{NWA-replacement} 
\end{equation}
which is motivated by the observation that the Breit-Wigner shape suppresses the 
contributions with $q^2\neq M^2$.  That is,
for small $\Gamma$, $\psi$ essentially eliminates all contributions of $f(q^2)$
except for the on-shell part, and in the limit 
$\Gamma\rightarrow 0$, $\psi(\sigma_r)$ and $\psi_\text{\,NWA}(\sigma_r)$ 
are asymptotically equal. 

However, for finite $\Gamma$ the $q^2$-dependence of the phase-space factors and 
residual matrix elements can cause a significant 
deformation of the shape of $D(q^2)\sigma_r(q^2)$ relative to the 
Breit-Wigner shape $D(q^2)$, resulting in a considerable shift 
of the maximum position and maximal value.
The effect is particularly strong when $M$ is close to the kinematic bounds, 
where threshold-type factors suppress the resonant contribution and shift the maximum.
Focusing on the impact of the $q^2$-dependence of the process-independent 
phase-space factors, we write 
$\psi(\sigma_r)=\int \frac{dq^2}{2\pi} D(q^2) \text{PS}(q^2)\tilde\sigma_r(q^2)$, 
where $\text{PS}(q^2)$ denotes the integrand factor arising from the 
phase-space element.
Even though the $q^2$-dependence of $\text{PS}(q^2)$ deforms the Breit-Wigner, 
the shape of $D(q^2)\text{PS}(q^2)$ is also strongly peaked and thus suggests 
a NWA-inspired approximation, which takes into account the 
shift of the maximum position caused by $\text{PS}(q^2)$.
More specifically, we propose to substitute the mass of the 
resonance $M$ with an effective mass $M_\text{eff}$ in $\psi_\text{\,NWA}(\sigma_r)$.
In analogy to $M^2$ being the maximum position of the Breit-Wigner, 
$M_\text{eff}^2$ is given by the position of the maximum of $D(q^2)\text{PS}(q^2)$:
\begin{gather}
\begin{split}
\psi_\text{\,NWA}(\sigma_r)\quad\to\quad &\psi_\text{\,PSINWA}(\sigma_r)=\frac{1}{2M_\text{eff}\Gamma}\int 
dq^2\delta(q^2-M_\text{eff}^2)\sigma_r(q^2)\\
&\quad\text{with}\ \;M_\text{eff}:=\left(\operatorname{arg}\max_{q^2} D(q^2)\text{PS}(q^2)\right)^{1/2}\,.
\end{split}
\label{NWA-replacement2} 
\end{gather}
The thus defined effective mass only exploits kinematic information and 
is hence universal for the class of processes with identical phase-space 
properties.\footnote{Since $M_\text{eff}^2$ can be obtained by numerical, 
one-dimensional maximization of $D(q^2) \text{PS}(q^2)$ with $q^2=M^2$ as 
suitable initial value, the computational complexity of its determination 
is negligible.  In theories with mass relations it may be necessary to 
adjust other parameters to maintain theoretical properties like gauge invariance.}
Due to $\lim_{\Gamma\rightarrow0}M_\text{eff}=M$, the deviation of the effective mass 
from the physical mass is negligible unless $M$ is close to the kinematic bounds.
Therefore, the proposed phase-space improved narrow-width approximation (PSINWA)
does not result in significant deviations in cases where the standard NWA gives 
$\calO(\Gamma/M)$-accurate results.
On the other hand, when $M$ approaches a kinematic bound the behavior is improved:
Since the distance between $M_\text{eff}$ and the bound stays finite
the PSINWA result does not vanish.  The PSINWA error is therefore bounded in contrast 
to the diverging standard NWA error.

\begin{figure}
\vspace*{0.cm}
\includegraphics[width=0.49\linewidth]{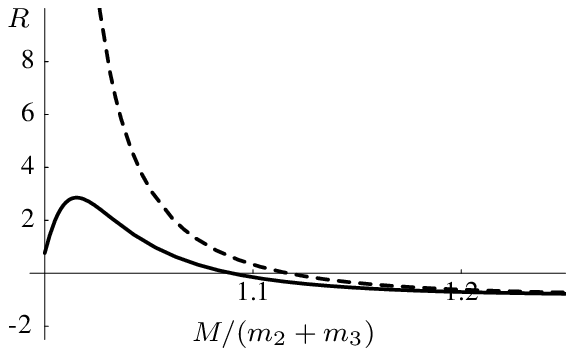}
\includegraphics[width=0.49\linewidth]{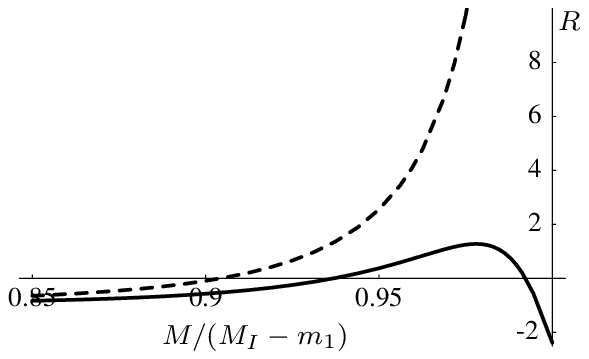}
\caption{\label{fig:deformation1}
Comparison of the phase-space improved (solid) and standard (dashed) NWA error for a 
scalar resonant $3$-body decay, i.e.~process SSS-SSS of Sec.~\ref{modindep},
when the resonance mass $M$ approaches the kinematic bounds $m_2+m_3$ (left)
and $M_I-m_1$ (right) (see Fig.~\ref{threebody-kinematics}).
Displayed is the relative approximation error in units of resonance width/mass, 
i.e.~$R=(\Gamma_\text{OFS}/\Gamma_\text{[PSI]NWA}-1)/(\Gamma/M)$, for 
$\Gamma/M=0.05$, $m_1=m_2=0$ and $m_3/M_I=0.2$.}
\end{figure}

To exemplify this method, we consider the scalar resonant $3$-body decay SSS-SSS 
(see Sec.~\ref{modindep}).  With $\beta(m,M):=\sqrt{1-m^2/M^2}$ and kinematic 
conventions as in Fig.~\ref{threebody-kinematics}, the 3-particle phase-space 
element is given by
\[ d\phi= d\phi_p \frac{dq^2}{2\pi} d\phi_d\,, \]
where
\begin{eqnarray*}
d\phi_p&=&\frac{1}{32\pi^2}\beta(\sqrt{q^2}+m_1,M_I)\beta(\sqrt{q^2}-m_1,M_I)d\Omega_p\,,\\
d\phi_d&=&\frac{1}{32\pi^2}\beta(m_2+m_3,\sqrt{q^2})\beta(m_2-m_3,\sqrt{q^2})d\Omega_d\,.
\end{eqnarray*}
$M_\text{eff}^2$ is thus obtained via maximization of 
\begin{equation*}
\text{PS}(q^2)D(q^2)\propto\frac{\beta(\sqrt{q^2}+m_1,M_I)\beta(\sqrt{q^2}-m_1,M_I)\beta(m_2+m_3,\sqrt{q^2})\beta(m_2-m_3,\sqrt{q^2})}{(q^2-M^2)^2+M^2\Gamma^2}\,.
\end{equation*}
In Fig.~\ref{fig:deformation1}, we compare the error of the phase-space improved 
and standard NWA when the kinematic bounds are approached.
It shows that the effective-mass method mitigates the error $|R|$ to 
less than $3\Gamma/M$ for $M$ arbitrarily close to the 
kinematic bounds.  The divergence of the standard NWA error in this region 
is also clearly visible.  With increasing distance to the kinematic bounds,
on the other hand, the PSINWA and standard NWA results converge and are both 
of $\calO(\Gamma/M)$.

The PSINWA is general in the sense that the effective mass does not depend 
on the process matrix elements, and as effective-mass method it preserves 
the on-shell simplifications of the standard NWA.
The former strength is, however, also a potential shortcoming,
since for some processes Breit-Wigner deformation is also caused by the 
$q^2$-dependence of the residual matrix elements, which may have to be taken 
into account to achieve the desired approximation accuracy.  As long as the 
deformed Breit-Wigner shape is still strongly peaked, the proposed method 
can be extended: a process-specific effective mass can be calculated by maximizing 
$f(q^2):=D(q^2)\text{PS}(q^2)|\mathcal M(q^2)|^2$ (and possibly also treating the
width as free parameter).\footnote{For more complicated phase spaces and/or matrix 
elements the function to maximize, namely $f(q^2)$, may not be available
in analytical form.  The computation of $M_\text{eff}^2$ will then be 
more expensive.}
For cases where the Breit-Wigner deformation is so severe that a modified mass
is not sufficient to obtain the desired accuracy, it may be possible to 
successfully apply the process-specific method proposed in 
Ref.~\cite{Kauer:2007nt}.


\section{Summary\label{conclusions}}
We studied the general properties of the NWA with polarization/spin decorrelation.  
After defining the NWA and clarifying the treatment of polarization and spin 
correlations, we proved for sufficiently inclusive rates of arbitrary 
resonant decay and scattering processes with an on-shell intermediate state 
decaying via a cubic or quartic vertex
that decorrelation effects vanish and the NWA is 
of order $\Gamma$.
When applied in perturbative calculations, the NWA uncertainty is commonly 
estimated as between $\approx\Gamma/M/3$ and $\approx 3\Gamma/M$.
We tested this assumption by systematically 
determining the NWA accuracy numerically for all 
resonant $3$-body decays involving scalars, spin-$\frac{1}{2}$ fermions or vector 
bosons.  
We found that the approximation error will exceed the $\calO(\Gamma/M)$ estimate for 
mass configurations in an extended vicinity of segment kinematic bounds. 
This is due to a significant distortion of the Breit-Wigner peak and tail, 
which is effected by the $q^2$-dependence of the phase-space elements and 
the residual matrix elements.
While factorizing coupling parameters do not affect the relative NWA error,
for processes with SFF or VFF vertices a strong dependence 
on the relative strength of the chiral couplings can exist.
We specialized the general results to MSSM benchmark scenarios and presented 
results for SPS 1a.  We found that significant off-shell 
corrections can occur -- similar in size to QCD corrections.  
To simplify a combined treatment we proposed a modified, process-independent 
approximation that exhibits an improved accuracy close to 
kinematic boundaries while preserving the simplifying power of the 
standard NWA.


\begin{acknowledgments}
We thank the referee for bringing Ref.~\cite{Dicus:1984fu} to our attention.
This work was supported by the BMBF, Germany (contract 05HT1WWA2).
\end{acknowledgments}

\end{document}